\documentstyle[emulateapj]{article}
\slugcomment{The Astrophysical Journal, 586:000-000, 2003, March 20}

\def\apj{ApJ}

\def\aj{AJ}
\def\aap{A\&\hskip-1pt A}

\def\mnras{MNRAS}
\def\pasp{PASP}

\def\nat{Nature}

\newcommand{\vect}[1]{{\mbox{\boldmath $#1$}}}

\def\thetae{\theta_{\rm E}}
\def\te{t_{\rm E}}
\def\dt{\Delta t}
\def\dos{D_{\rm os}}
\def\dls{D_{\rm ls}}
\def\dol{D_{\rm ol}}
\def\drel{D}
\def\murel{\mu_{\rm rel}}
\def\kpc{\rm kpc}
\def\msun{M_\odot}
\def\sdt{\sigma_{\dt}}
\def\ste{\sigma_{\te}}
\def\sphi{\sigma_{\phi}}
\def\sthe{\sigma_{\thetae}}

\def\muas{\mu{\rm as}}
\def\mas{{\rm mas}}
\def\bu{\mbox{\boldmath $u$}}
\def\bphi{\mbox{\boldmath $\varphi$}}
\def\bt{\mbox{\boldmath $\theta$}}

\def\dbtcl{\delta \bt_{cl}}

\def\kms{\rm km~s^{-1}}
\def\retilde{\tilde r_{\rm E}}

\def\eq#1{equation (\ref{#1})}

\begin{document}

\lefthead{GAUDI, GRAFF, \& HAN} \righthead{ANGULAR RADII OF STARS VIA MICROLENSING}

\title{Angular Radii of Stars via Microlensing}
 
\author{B. Scott Gaudi\altaffilmark{1}}
\affil{School of Natural Sciences, Institute for Advanced Study, Einstein Drive, Princeton, NJ 08540; gaudi@sns.ias.edu}
\author{David S.\ Graff\altaffilmark{2}}
\affil{Department of Astronomy, The Ohio State University, 140 West 18th Avenue, Columbus, OH 43210; graff@astronomy.ohio-state.edu}
\and
\author{Cheongho Han}
\affil{Department of Physics, Institute for Basic Science Research, Chungbuk National University, Chongju 361-763, Korea; cheongho@astroph.chungbuk.ac.kr}
\altaffiltext{1}{Hubble Fellow}
\altaffiltext{2}{Current address: Math and Science, United States Merchant Marine Academy,
Kings Point, New York 11024; dgraff@usmma.edu}

\begin{abstract}

We outline a method by which the angular radii of giant and main
sequence stars located in the Galactic Bulge can be measured to a few
percent accuracy.  The method combines comprehensive ground-based
photometry of caustic-crossing bulge microlensing events, with a
handful of precise ($\sim 10 \muas$) astrometric measurements of the
lensed star during the event, to measure the angular radius of the
source, $\theta_*$.  Dense photometric coverage of one caustic
crossing yields the crossing time scale $\dt$.  Less frequent coverage
of the entire event yields the Einstein timescale $\te$ and the angle
$\phi$ of source trajectory with respect to the caustic.  The
photometric light curve solution predicts the motion of the source
centroid up to an orientation on the sky and overall scale.  A few
precise astrometric measurements therefore yield $\thetae$, the
angular Einstein ring radius.  Then the angular radius of the source
is obtained by $\theta_*=\thetae(\dt/\te) \sin{\phi}$.  We argue that
the parameters $\te, \dt, \phi$, and $\thetae$, and therefore
$\theta_*$, should all be measurable to a few percent accuracy for
Galactic bulge giant stars using ground-based photometry from a
network of small (1m-class) telescopes, combined with astrometric
observations with a precision of $\sim 10\muas$ to measure $\thetae$.
We find that a factor of $\sim 50$ times fewer photons are required to
measure $\thetae$ to a given precision for binary-lens events than
single-lens events.  Adopting parameters appropriate to the {\it Space
Interferometry Mission} (SIM), we find that $\sim 7$ minutes of SIM
time is required to measure $\thetae$ to $\sim 5\%$ accuracy for giant
sources in the bulge. For main-sequence sources, $\thetae$ can be
measured to $\sim 15\%$ accuracy in $\sim 1.4$ hours.  Thus, with access
to a network of 1m-class telescopes, combined with 10 hours of SIM
time, it should be possible to measure $\theta_*$ to $5\%$ for
$\sim$80 giant stars, or to $15\%$ for $\sim$7 main sequence stars.
We also discuss methods by which the distances and spectral types of
the source stars can be measured.  A byproduct of such a campaign is a
significant sample of precise binary-lens mass measurements.

\end{abstract}


\section{Introduction}

Although of fundamental importance to stellar astrophysics, precise
measurements of angular radii are generically difficult to acquire
routinely and in a model-independent way.  Classical direct methods of
measuring stellar radii include lunar occultations, interferometry,
and eclipsing binaries.  Lunar occultation measurements yield precise
angular radii (see Richichi et al.\ 1999 and references therein), but
the number of stars to which this technique can be applied is limited.
The number of direct measurements using interferometers has recently
increased dramatically with advent of, e.g.\ the Palomar Testbed
Interferometer (van Belle et al.\ 1999a, Colavita et al.\ 1999), 
and the Navy Prototype Optical Interferometer (Armstrong et al.\ 1998, Nordgren et al.\ 1999),
and is likely to continue to increase as more technologically advanced interferometers come online.
Unfortunately, both lunar occultation and interferometric angular
diameter measurements have traditionally been 
primarily limited to nearby, evolved stars.  Angular radii of
main-sequence stars can be determined using detached eclipsing
binaries (i.e.\ Popper 1998), however the large amount of data (both
photometric and spectroscopic) required to yield accurate radii
determinations makes this method prohibitive.  Thus, of the $\sim300$
direct, precise angular diameter measurements compiled by van Belle
(1999), the overwhelming majority, $\sim 85\%$, are of evolved stars.
Finally, it will be difficult to acquire a large
sample of angular radii determinations of stars with metallicity
considerably smaller than solar using these methods, due to the
paucity of metal-poor stars in the local neighborhood.

Here we present a method, based on a suggestion by Paczy\'nski (1998),
of measuring angular radii of stars that overcomes some of the
difficulties inherent in the classical methods.  This method employs
the extraordinary angular resolution provided by caustics in
gravitational microlensing events, and as such is yet another in the
growing list of applications of microlensing to the study of stellar
astrophysics (see Gould 2001 for a review).  The original suggestion of
Paczy\'nski (1998) was to invert the method of Gould (1994) for
measuring the relative source-lens proper motion $\murel$ in
microlensing events.  If the lens transits the source in a
microlensing event, precise photometry can be used to determine the
time it takes for the lens to transit one source radius,
$t_*=\theta_*/\murel$, where $\theta_*$ is angular radius of the
source.  An estimate of $\theta_*$, using an empirical color-surface
brightness relation, together with a measurement of the flux of the
source, can then be used to estimate $\murel$, which Gould (1994)
argued could be used to constrain the location of the lens.  However, 
as Paczy\'nski (1998) pointed out, it is possible to independently measure 
the angular Einstein ring radius of the lens,
\begin{equation}
\thetae=\sqrt{{{4 G M}\over \drel c^2} },
\label{eqn:thetae}
\end{equation}
by making precise astrometric measurements of the centroid shift of
the source during the microlensing event using, i.e., the {\it Space
Interferometry Mission} (SIM).\footnote{http://sim.jpl.nasa.gov} Here
$M$ is the mass of the lens, $\drel$ is defined by, $\drel\equiv
\dos\dol/\dls$, and $\dos$, $\dol$, and $\dls$ are the distances
between the observer-source, observer-lens, and lens-source,
respectively.  Since $\murel = \thetae/\te$, by combining the
measurement of $\thetae$ with the Einstein timescale $\te$ of the
event determined from the light curve, it is possible to measure
$\theta_*$ for the source stars of microlensing events.
We show that, with reasonable expenditure of
resources, it should be possible to measure angular radii of a
significant sample ($\sim 80$) of giant stars in
the bulge to an accuracy of $\la 5\%$, or $\sim 7$ main-sequence
stars to an accuracy of $\la 15\%$.   Limb-darkening determinations
should also be possible for the majority of the sources, and most will
be relatively metal poor as compared to those for which angular radii
determinations are currently available.

Although measurements of $\theta_*$ can be made using single-lens events,
in \S\ref{sec:bvs} we argue that
this method is better suited to caustic-crossing binary-lens events, 
which are more common, easier to
plan for, and considerably less resource-intensive than source-crossing single-lens events.  
We describe in
some detail the basic method of measuring $\theta_*$ for the source
stars of caustic-crossing binary-lens events in 
\S\ref{sec:method}, including a discussion of the expected errors on
the individual parameters that enter into the measurement.  We discuss
various subtleties, complications, and extensions to the method in
\S\ref{sec:discussion}, and also present an estimate of the number of
$\theta_*$ measurements that might be made in this way.  Finally, we summarize and
conclude in \S\ref{sec:conclusion}.

\section{Binary versus Single Lens Events}\label{sec:bvs}

The primary requirement to be able to measure $\thetae$ in a
microlensing event is that the source should be resolved by the gravitational
lens.  This effectively means that the source must cross a caustic in
the source plane.  Caustics are the set of positions in the source
plane where the determinant of the Jacobian of the lens mapping from
source to lens plane vanishes, and where the magnification therefore
is formally infinite.  Large gradients (with respect to source
position) in the magnification exist near caustics, enabling the
resolution of the source.  Generically, microlenses come in two
classes: single and binary lenses.  Here `binary lens' means a lens
systems composed of two masses with angular separation of order
the angular Einstein ring radius of the system.  
Very close and very wide binaries act essentially as single lenses. 
Single lenses have a caustic that consists of a single point at the position of the lens. 
In these cases, the magnification close to the caustic diverges as the inverse
of the distance to the caustic.  In contrast, the caustics of binary
lenses are extended, and can cover a significant fraction of the
Einstein ring.  The caustics of binary lenses generically consist of
two types of singularities, folds and cusps.\footnote{For binary-lenses,
higher-order, beak-to-beak singularities can exist for specific combinations
of the binary-lens mass ratio and angular separation in units of $\thetae$.  However, 
folds and cusps are the only stable singularities of any lens system 
(Petters, Levine, \& Wambsganss 2000).  Beak-to-beak singularities are
unstable in the sense that, 
for infinitesimally small
variations in the lens parameters, a beak-to-beak singularity disintegrates
into two cusp-type singularities.  Thus the set of parameters where
beak-to-beak singularities are expected is formally sparse, and practically 
small.  Interestingly, Alcock et~al.\ (2000a) suggested that an observed event
may have been due to a source crossing a beak-to-beak singularity in a binary-lens, although 
Alcock et~al.\ (2000b) seem to favor the interpretation that this event is due
to a background supernova.} Near a fold, 
the caustic is well-described by generic linear fold
singularities, for which the magnification locally diverges inversely
as the square-root of the distance to the caustic (Schneider \& Weiss 1986,
Schneider, Ehlers, \& Falco 1992, Gaudi \& Petters 2002a).  Cusps are points where two
fold caustics meet, and the magnification for cusps locally diverges
roughly inversely as the distance to the cusp point, similar
to the magnification pattern near the point caustic of a single lens
(Schneider \& Weiss 1992, Schneider, Ehlers, \& Falco 1992, Gaudi \& Petters 2002b).  
The fact that the magnification near the point caustic of a
single lens or near a cusp diverges inversely as the distance to the
singularity, rather than as the square root of the distance to the
singularity as with folds, means that for a given source size, the
`resolving power' of fold caustic crossings is less than single lens
or cusp crossings.

Although single-lens events are more well suited to studies
of stellar atmospheres, they are much less useful
for measuring the sizes of stars for three main reasons.
They are generally rarer than fold caustic crossings, their
crossings cannot be predicted in advance, and their centroid 
motion is more difficult to measure.

Caustic-crossing binaries comprise roughly $f_{cc}=7\%$ of all
events toward the Galactic bulge (Alcock et~al.\ 2000a; Udalski et~al.\ 2000).  Of the
remaining $1-f_{cc}$ events, which we will conservatively assume due
to single lenses, only a fraction $\theta_*/\thetae$ will exhibit
caustic crossings.  Therefore, the expected ratio of binary-to-single
events for which the source is resolved (and thus measurement of
$\theta_*$ is possible) is
\begin{equation}
\Gamma_{b/s}\simeq\frac{\thetae}{\theta_*} \frac{f_{cc}}{1-f_{cc}}
\sim 4 \left(R_* \over 10 R_\odot\right)^{-1},
\label{eqn:fbs}
\end{equation}
where $R_*$ is the physical radius of the source, 
and we adopted $\dol=6~\kpc$, $\dos=8~\kpc$, and $M=0.3\msun$ for
the scaling relation on the extreme right hand side.  Although not
overwhelming for giant sources, for main sequence sources we expect at
least an order of magnitude more binary lensing events for which the
source is resolved.  The ratio of the number of fold-to-cusp crossings
is roughly
\begin{equation}
\Gamma_{f/c}\simeq\frac{\thetae}{\theta_*} N_{cusp}^{-1} \sim 55
N_{cusp}^{-1} \left(R_* \over 10 R_\odot\right)^{-1},
\label{eqn:ffc}
\end{equation}
where $N_{cusp}$ is the number of cusps, which is either 6, 8, or 10
for a binary lens, depending primarily on $d$.  Thus the overwhelming
majority of events for which it will be possible to measure $\theta_*$
will be fold caustic crossing binary-lens events.

Binary-lens fold caustic-crossing events also have the advantage that the 
second caustic crossing
can be predicted in advance.  This is because fold caustic crossings always
come in pairs, and it is typically easy to tell, even with sparse
sampling, that the first caustic crossing has occurred.  Then more
frequent sampling can be used to monitor the rise to the second
caustic, in principle enabling the prediction of the time of the
second crossing a day or more in advance (Jaroszy{\' n}ski \& Mao
2001), and allowing the marshaling of the resources necessary to
obtain the dense coverage of the second crossing needed to measure the
crossing time $\dt$ of the source (see \S\ref{sec:dt}).  In contrast,
a single lens caustic crossing can only reliably be `predicted' at
about the time it begins.

Finally, it is considerably harder to measure $\thetae$ 
for caustic-crossing single-lens events than binary-lensing
events.  Single lens events have a maximum absolute centroid shift relative
to the unlensed source position of $\thetae/8^{1/2}$, whereas 
binary lensing events can exhibit large variations of size $\sim \thetae$ or more
when the source crosses the caustic (Han, Chung, \& Chang 1999).  
Therefore 
considerably more time will generally be required to determine $\thetae$ to
given accuracy for single-lens events than for binary-lens events.
Since the astrometric measurements essentially require the
capabilities of SIM (or some similarly precious instrument), it is highly desirable to minimize the
amount of time spent on this step.

Given the above arguments, we conclude that fold caustic crossing
binary lensing events are the most suitable for use in routinely
measuring $\theta_*$.  We will therefore focus on this case for the
remainder of the discussion, however we will briefly revisit
single-lens and cusp-crossing events in \S\ref{sec:cusp}.

\begin{figure*}
\epsscale{2.2}
\centerline{\plotone{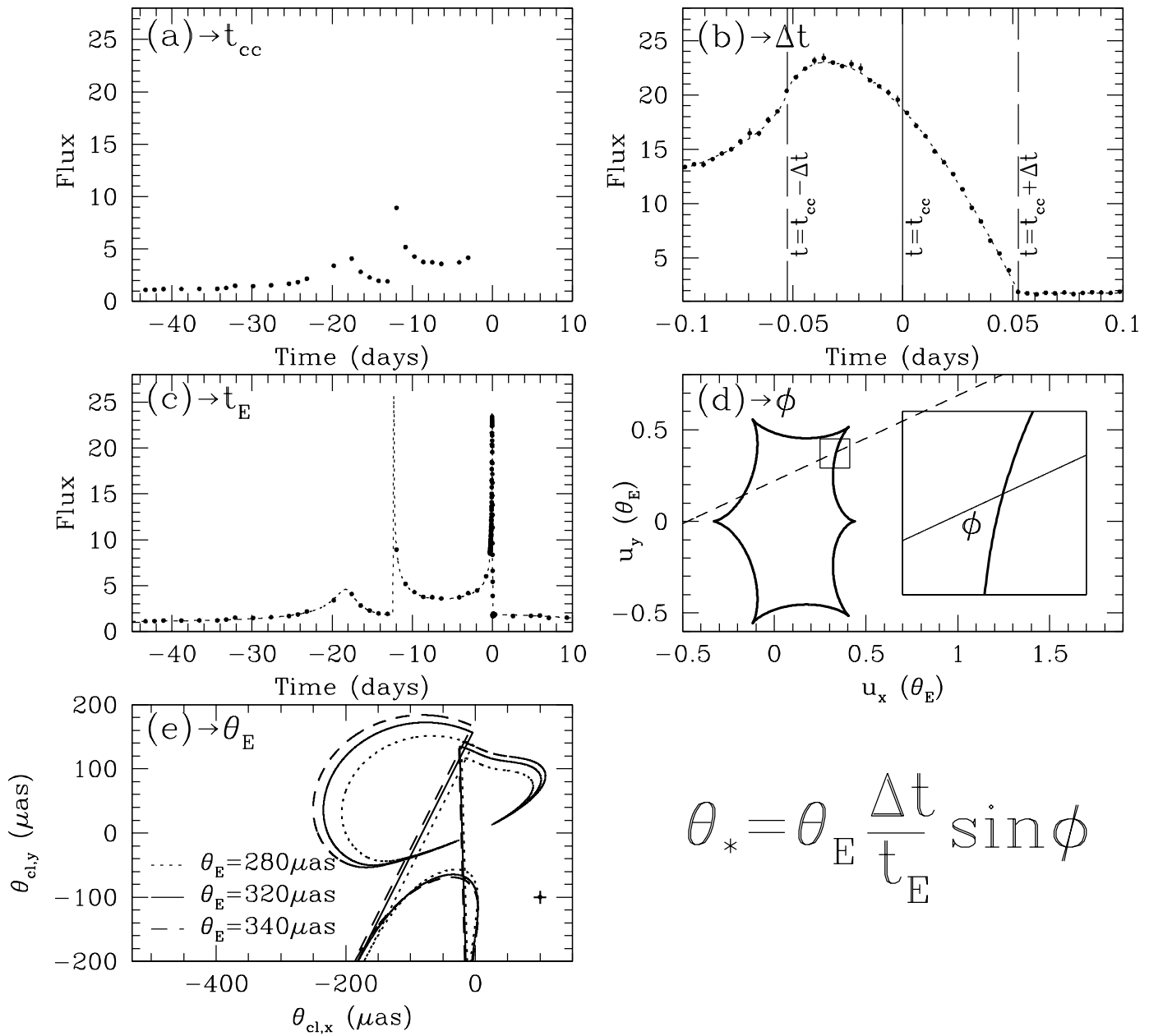}}
\caption{ Schematic illustration of the various steps involved in measuring
the angular radius of the source star of a caustic-crossing
binary-microlensing event.  (a) Infrequent ($\sim 1$ per day)
observations of bulge microlensing events from a single site reveal a
caustic-crossing event real-time, enabling a prediction for, and
intense photometric monitoring of, the second caustic crossing.  (b)
Fitting the photometry near the second crossing to a generic fold
caustic model yields the caustic crossing timescale $\dt$.  (c,d)
Fitting to the entire photometric dataset to a binary-lens model
yields the Einstein timescale $\te$ of the event, as well as the angle
$\phi$ of the trajectory with respect to the caustic.  (e) The
photometric solution predicts the motion of the centroid of the source
up to an unknown scale and orientation on the sky.  A few, precise
astrometric measurements can then be used to determine $\thetae$, the
angular Einstein ring radius.  The cross shows errorbars of $10\muas$.
The angular source size is then given
by $\theta_*=\thetae(\dt/\te)\sin\phi$. }
\end{figure*}

\section{The Method}\label{sec:method}

Consider a binary-lens event in which the source crosses a simple
linear caustic.\footnote{A caustic can generally be approximated as a simple
linear fold when the curvature of the caustic is everywhere small on angular
scales of ${\cal O}(\theta_*)$, and when the angle of incidence 
of the source trajectory  to the caustic is not small.  This approximation will break down
when the source is large compared to the overall size of the caustic,
when the source crosses near a cusp, or when the source
`straddles' the caustic for a long time due to a small incidence
angle.  Although it will still be
possible to measure $\theta_*$ for such events, the relation between
the observables and $\theta_*$ is less straightforward.  We discuss
such cases in \S\ref{sec:cusp}.}  Defining
$\dt$ as the (one-half) the time it takes for the source to
completely traverse the caustic, and $\phi$ as the angle between the
source trajectory and the tangent to the caustic at the crossing
point, then the time for the lens to cross the angular radius of the
source is $t_*=\dt\sin\phi$.  However, we also have that $t_* =
\theta_*/\murel$, where again $\murel$ is the relative source-lens
proper motion.  Combining these expressions, we have that
$\theta_*=\murel \dt\sin\phi $.  Using the definition of $\murel$, we
can write the angular source radius $\theta_*$ as the following
function of observables,
\begin{equation}
\theta_* = \thetae \frac{\dt}{\te} \sin{\phi}.
\label{eqn:thetas}
\end{equation}
The process of $\theta_*$ measurement can therefore be subdivided into three
basic steps:
\begin{description}
\item[{(1)}] Measurement of the caustic crossing timescale $\dt$ from
a single photometrically well-resolved caustic crossing.
\item[{(2)}] Measurement of the angle $\phi$ and timescale $\te$ from
the global fit to the binary-lens light curve.
\item[{(3)}] Measurement of the angular Einstein ring radius $\thetae$
using precise astrometric measurements of the source centroid.
\end{description}
Each of these steps are illustrated schematically in Figure 1.

In the following subsections, we consider each of these steps in more
detail.  We outline the basic requirements for the measurement of each
of the four parameters ($\dt,\phi,\te$, and $\thetae$), and the
expected accuracy with which each can be determined assuming
reasonable expenditure of observing resources.

\subsection{Measuring $\dt$}\label{sec:dt}

When the source is interior to a binary-lens caustic, five images are
created. As the source approaches a fold caustic, two of these
images brighten and merge in a characteristic way, and
eventually disappear when the source completely exits the caustic.  
In contrast to the significant
brightening of the two images associated with the fold, the remaining
three images generally vary only slowly over the timescale of the
crossing.  All fold caustics locally have this generic behavior, and
the magnification $A(t)$ of the source near a caustic crossing as a
function of time is typically well-fitted by the functional form
(Albrow et~al.\ 1999b),
\begin{equation}
A(t)=\left(\frac{t_r}{\dt}\right) G_0\left(\frac{t-t_{cc}}{\dt}\right)+A_{cc} +\omega(t-t_{cc}),
\label{eqn:foft}
\end{equation}
where $t_r$ is the effective rise time of the caustic, which is
related to the local derivatives of the lens mapping (Petters et~al.\
2001, Gaudi \& Petters 2002), $t_{cc}$ is the time when the center of
the source crosses the caustic, $A_{cc}$ is the magnification of all
the images unrelated to the fold caustic at $t=t_{cc}$, $\omega$ is
the slope of the magnification of these images as a function of time,
and $G_0(x)$ can be expressed in terms of complete elliptic integrals of
the first and second kind (Schneider \& Weiss 1987).
Note that \eq{eqn:foft} is only formally appropriate for a simple
linear fold caustic.
 Thus, for a well-sampled fold caustic crossing, $\dt$ can be determined
essentially independently of the global geometry of the event, and
indeed without reference to the photometric data away from the
crossing itself.  
In practice, the magnification
is not directly observable, but rather the flux $F(t)$ as a function
of time.  The form for $F(t)$ takes on a similar form as
\eq{eqn:foft}, but with a slightly different parameterization (see
Albrow et~al.\ 1999b).  

Equation (\ref{eqn:foft}) assumes a uniform source.  This will likely
be a poor approximation in optical bands, and assuming uniform source
in the presence of limb-darkening may result in a systematic
underestimate of $\dt$, and therefore $\theta_*$ since, the effect of
limb-darkening can partially compensated for by a smaller
(dimensionless) source size, at least for poorly-sampled light curves.
However, for well-sampled caustic crossings and $\sim 1\%$ photometric
accuracy, it should generally be possible to accurately measure both
the source size and limb-darkening coefficient(s) (Rhie \& Bennett
1999).  Such data quality is readily achievable; indeed, independent
limb-darkening and (dimensionless) source size measurements have been
made for the source stars of at least five microlensing events (Albrow
et~al.\ 1999a, Afonso et~al.\ 2000, Albrow et~al.\ 2000, Albrow
et~al.\ 2001, An et~al.\ 2002).  A generalized form of \eq{eqn:foft}
that includes simple limb-darkening can be found in 
Albrow et~al.\ (1999b) and Afonso et~al.\ (2001).

By fitting \eq{eqn:foft} (or a generalized form of it) 
to the light curve near a well-sampled fold
caustic crossing, one can derive the parameters $t_r, t_{cc}, A_{cc},
\omega$, and $\dt$.  The parameters $t_{cc}, A_{cc}$, and $t_r$ can
subsequently be used to constrain the global solution to the entire
light curve (see Albrow et~al.\ 1999b).
However, of primary interest here is the parameter $\dt$, whose value
is essentially independent of the global solution.  See Figure 1(b).
This means that $\dt$ and the parameters determined from the global
solution, $\te$ and $\phi$, will be essentially uncorrelated.

In order to be able to determine $\dt$ from the caustic crossing data
alone, the caustic crossing must be well-sampled, so that the
parameters in \eq{eqn:foft} be well-constrained.  Practically, this
requires forewarning of the caustic crossing.  Fortunately, this is
generally possible with only photometry available from the
collaborations which survey the Galactic bulge and find the alert the
microlensing events real-time (EROS, Afonso et~al.\ 2001; MOA, Bond
et~al.\ 2001; OGLE, Udalski et~al.\ 2000, Woziak et~al.\ 2001),
although improved predictions would be possible with continuous
photometry (Jaroszy{\' n}ski \& Mao 2001). Thus no additional
resources need to be invested to predict caustic crossings.  However, as
we discuss in \S\ref{sec:teandphi}, some additional sampling of the
overall light curve may be needed to constrain $\te$ and $\phi$ and
determine a unique global solution.

The accuracy with which $\dt$ can be determined for a given
caustic crossing will depend not only on the intrinsic parameters of
the caustic crossing, but also on the sampling rate and photometric
accuracy near the caustic crossing, which in turn will depend on
weather, blending, etc., which tend to vary in a stochastic manner.
Therefore it is not very useful to attempt to quantify the expected
errors on $\dt$ for an idealized observing setup.  However we can
obtain an order-of-magnitude estimate for $\sdt$ by examining
measurements of $\dt$ from published analyses of observed
caustic-crossing events.  We discuss these determinations more
thoroughly in \S\ref{sec:errors}.   Typically, well-covered caustic
crossings yield fractional errors of $\sdt/\dt \sim 1\%$.

\subsection{Measuring $\te$ and $\phi$}\label{sec:teandphi}

The parameters $\te$ and $\phi$ are determined from the global
solution to the overall geometry of the binary-lens light curve.  The
relationship between the salient features of an observed light curve,
and the canonical parameters of a binary-lensing event, are generally
not obvious or straightforward.  This fact generally makes fitting an
observed light curve, and thus inferring the parameters $\te$ and
$\phi$, quite difficult (see Albrow et~al.\ 1999b for a thorough
discussion).  Although many methods have been proposed to overcome
these difficulties, the lack of obvious correspondence between these
parameters of interest and the light curve morphology generally
implies that it is difficult to make general statements about the
kinds of observations that are needed to reliably measure $\te$ and
$\phi$.  However, we can make some generic comments.  For
caustic-crossing binary lenses, one potential observable is the time
$\delta t_{cc}$ between caustic crossings.  Between caustic crossings,
the magnification is typically considerably larger than outside the
caustic, and therefore even with sparse ($\sim 1$ day) sampling, it
should be possible to determine to reasonable precision the time of
the first caustic crossing, provided that data exist before the first
crossing.  The requisite photometry will generally be acquired by the
survey collaboration(s).  Once the binary-lens event is alerted, more
frequent photometry can be acquired by follow-up collaborations
(PLANET, Albrow et~al.\ 1998; MPS, Rhie et~al.\ 2000), thus
mapping the ``U''-shaped curve between the caustic crossings.  This
shape, combined with information from cusp-approaches (or lack thereof)
just outside the caustic, provides information about the shape of the
caustic and the trajectory of the source through it.  From this, the
angle $\phi$ of the trajectory with respect to the caustic can be
derived, and also the distance $d_{cc}$ between the caustic crossings
in units of $\thetae$.  Then, the timescale is given by
$\te=t_{cc}/d_{cc}$.  

Although the above analysis is highly trivialized, it
does suggest that the following steps should be taken to ensure an
accurate measurement of $\te$ and $\phi$.  First, it is important to
constrain the time of both caustic crossings to reasonable precision.
This means that the survey collaborations should sample on timescales
no less than a few days.  Second, followup photometry should be
initiated relatively soon after the first crossing, to measure the
shape of the intra-caustic light curve reasonably well.  This is also
necessary in order to predict the second caustic crossing (Jaroszy{\' n}ski \& Mao 2001).
Furthermore, the follow-up photometry should continue past the second
caustic crossing, to detect cusp-approaches or cusp-crossings (or the lack thereof).

Due to the difficulties inherent in fitting binary-lens light curves,
it would be extremely difficult to attempt to predict the expected
errors on $\te$ and $\phi$ for a hypothetical observing scenario.
Furthermore, due to the complicated relation between observables and
parameters implies that these errors are likely to depend strongly on
the geometry of the event and light curve coverage, and therefore such `predictions' would
not be very useful.  However, we would like to
have an order-of-magnitude estimate for the expected errors given
reasonable light curve coverage.  From determinations of these
parameters in published caustic-crossing events, we can expect
fractional accuracies of a few percent, provided that the light curve is
well-covered, in the sense outlined above.  However, if the light curve
coverage is incomplete, then errors of $\ga 20\%$ are expected.
See \S\ref{sec:errors}.

\begin{table*}[t]
\begin{center}
\begin{tabular}{cccccccc}
\tableline
\multicolumn{1}{c}{Event Name} &
\multicolumn{1}{c}{$\sdt/\dt$} &
\multicolumn{1}{c}{$\sphi/\phi~(\phi)$} &
\multicolumn{1}{c}{$\sigma_{\rho_*}/\rho_*$} &
\multicolumn{1}{c}{$\ste/\te$} &
\multicolumn{1}{c}{$I$} &
\multicolumn{1}{c}{Ref.} & 
\multicolumn{1}{c}{Comments}\\
\tableline
\tableline
MACHO 98-SMC-1   & $0.9\%$ & $32\%$  ($36^\circ\hskip-3.5pt.9$) & ... & $23\%$ &  22.1 & 1,2 &Fold crossing \\
OGLE-1999-BUL-23 & $0.5\%$ & $0.2\%$  ($56^\circ\hskip-3.5pt.1$)  & ... & $1\%$  & 18.1 & 3 & Fold crossing\\
\hline
MACHO 95-BLG-30  & ...      & ...                             & $0.1\%$    & ...  & 13.4 & 4 &Single lens \\
MACHO 97-BLG-28  & ...      & ...                             & $0.3\%$ & ...  & 15.6 & 5 & Cusp crossing\\
MACHO 97-BLG-41  & ...      & ...                             & $5\%$ & ...&  16.8 & 6 & Rotating binary\\
EROS 2000-BLG-5  & ...	   & ...                             & $0.8\%$    & ...  & 16.6 & 7 & Parallax effects\\
\tableline
\tableline
\end{tabular}
\end{center}
\tablenum{1} {\bf Table 1} Fractional Errors on Parameters from Observed Events. \\
\tablerefs{ (1) Albrow et~al.\ (1999b), (2) Afonso et~al.\ (2000), (3)
Albrow et~al.\ (2001a), (4) Alcock et~al.(1997), (5) Albrow
et~al.(1999a), (6) Albrow et~al.\ (2000), (7) An et~al.(2002) }
\label{tbl:table1}
\end{table*}

\subsection{Measuring $\thetae$}\label{sec:thetae}

A global solution to the entire light curve effectively requires the specification of the 
vector position of the source
$\bu(t)$ as a function of time in units of $\thetae$, and the topology
of the lens, i.e.\ the mass ratio $q$ and projected separation $d$ in
units of $\thetae$.  These parameters yield not only 
the total magnification $A$ of all the images as a function of time.
but also the individual image positions $\bphi_i$ and magnifications $A_i$ as a function of time, 
Therefore,
it is also possible to predict the centroid $\bphi_{cl}$ of all the
individual microimages,
\begin{equation}
\bphi_{cl}= \frac{\sum_i A_i \bphi_i}{A}.
\label{eqn:btcl}
\end{equation}
Note that $\bphi_{cl}$ is the centroid shift with respect to the lens
position.  Is it customary to consider
the centroid shift with respect to the unlensed source
position, $\delta \bphi_{cl}\equiv \bphi_{cl}- \bu$.  The unlensed
source position $\bu$, 
which is comprised of the parallax and proper motion
of the source in some astrometric frame, can be determined via measurements of the
unlensed motion of the source. Since the astrometric effects falls off
very slowly (as $u^{-1}$), these must be obtained many $\te$ after the
event is over.  In fact, such measurements are not strictly needed in
order to measure $\thetae$ (but may be desirable for other reasons,
see \S\ref{sec:dos}).  Rather, one can simultaneously fit astrometric
measurements during the course of the event for the relative position 
and proper motion of the source (in order to establish a local astrometric reference frame), and 
the offset induced by microlensing.

The centroid shift $\bphi_{cl}$ is in units of $\thetae$, and its components are oriented
with respect to the projected binary-lens axis, whose orientation
$\alpha$ on the sky is unknown.  The observable centroid is
\begin{equation}
\bt_{cl}(t) \equiv \thetae 
\left(
\begin{array}{cc}
\cos\alpha&\sin\alpha\\ \\
-\sin\alpha&\cos\alpha
\end{array}
\right)
\bphi_{cl}(t).
\label{eqn:dbt}
\end{equation}
Alternatively, one can combine $\alpha$ and $\thetae$ and simply consider
the vector $\vect{\theta}_{\rm E}$. 
Thus the global solution to the photometric light curve yields not only
$\te$ and $\phi$, but also yields a prediction for the astrometric
curve, up to an unknown orientation and scale
$\thetae$.\footnote{Technically, this is only true for an astrometric
observer that is spatially coincident with the photometric observer.
This will not be true for astrometric observations with SIM which
will be in an Earth-trailing orbit.  We
discuss this complication in \S\ref{sec:parallax}.}  Therefore, by making a series of astrometric
measurements $\bt_{cl}(t)$ at several different times during the course of the event,
one can determine $\thetae$ and $\alpha$ via \eq{eqn:dbt}, and using the predictions for $\bphi_{cl}(t)$ from
the photometric solution.

The accuracy with which $\thetae$ can determined will depend on the geometry of
the event, the time of the astrometric measurements, and the time span between
the measurements.  Furthermore, as we discuss, there are, in reality, 
\centerline{{\vbox{\epsfxsize=9.0cm\epsfbox{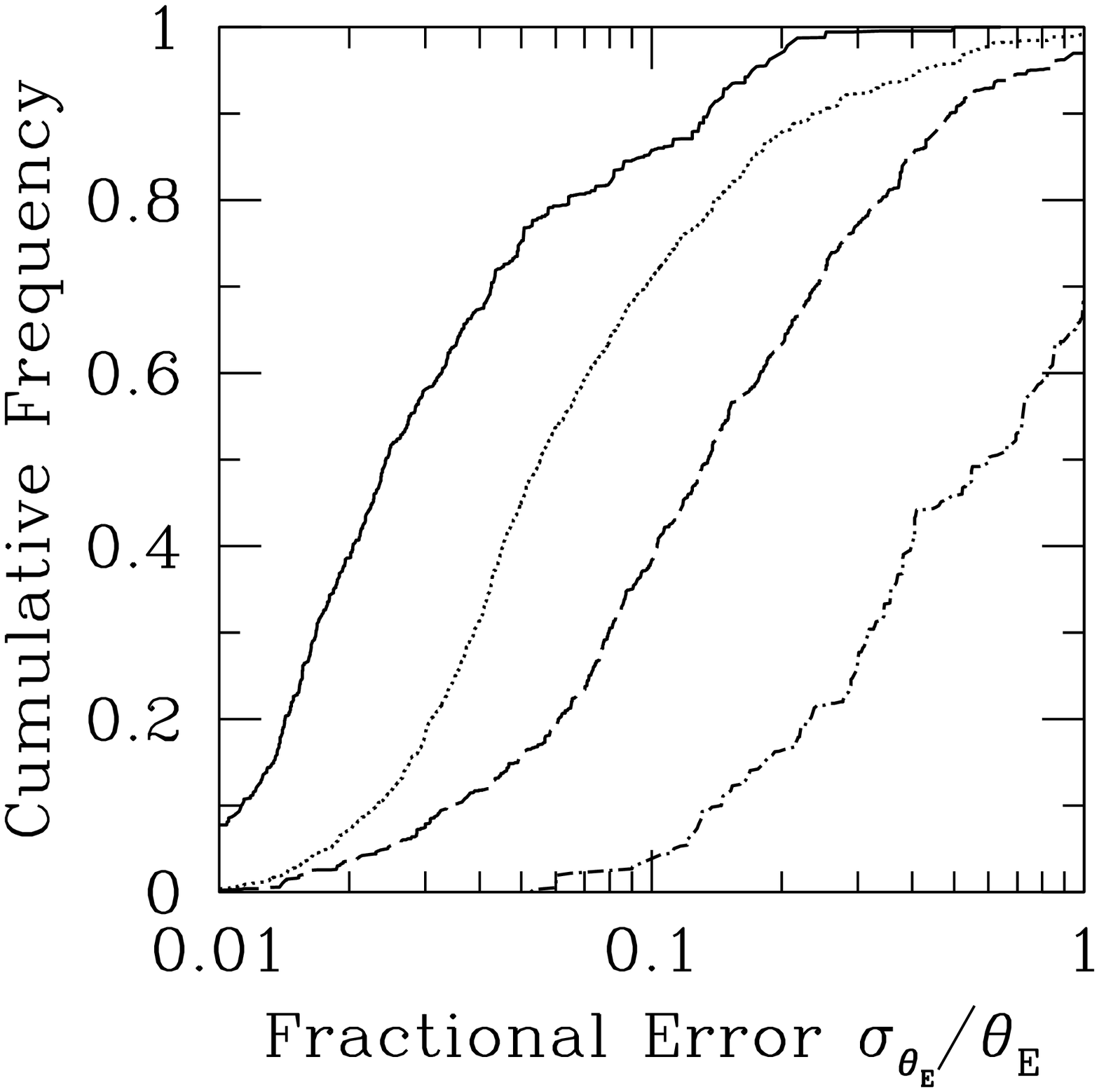}}}}
{\footnotesize { FIG. 2.-- The cumulative distribution of expected errors in $\theta_E$ in a
Monte Carlo simulation, assuming a contribution of blended
light of $0\%$ (no blending, solid), $1\%$ (dotted), $10\%$ (long dashed), and $90\%$ (dot-dashed).  See text. }
}\smallskip\\
additional parameters that must be determined from the astrometric data.  We therefore perform a Monte Carlo
simulation to estimate the accuracy with which $\thetae$ can be recovered. 
We explore whether the various parameters can be measured
independently, or are degenerate with the measurement of $\thetae$, and 
study the range of fractional uncertainties in measuring $\thetae$ for a
full ensemble of binary lenses.

Our simulation consists of the following elements: a
Galactic model and model of the ensemble of lenses to generate an
ensemble of microlensing events, an observational strategy, and a fit
of these observations to a set of parameters including $\thetae$.

Our ensemble of lenses is identical to that of Graff \& Gould (2002).
Briefly, we draw sources and lenses from self-lensing isothermal
sphere 8 kpc from the observer, and with two-dimensional velocity
dispersion of $220 \kms$.  Both masses in the binary lens are chosen
from the remnant mass function of Gould (2000).  We pick a flat
distribution in $\log(d)$, the logarithm of the dimensionless binary
separation.  The path of the source through the lens geometry is
chosen randomly with a uniform distribution of angular impact parameter
$b \thetae$, and we only consider paths that cross
caustics.\footnote{In contrast to the usual technique for single lenses
in which events are chosen from a uniform distribution in $b$, but are
weighted towards large $\thetae$ events by multiplying the mass
function by $M^{1/2}$.}

As discussed in Graff \& Gould (2002), in this ensemble of events
there are many more events with a short time between caustic
crossings, the {\it  caustic interior time $t_{\rm int}$}, than is
observationally detected by the MACHO and OGLE observing groups.  This
led these authors to suggest that most events with a short caustic
interior time are not detected as caustic crossing binaries, and to
define a caustic crossing detection efficiency ${\cal{E}}_b(t_{\rm
int})$ akin to the standard single lens detection efficiency
${\cal{E}}_s(t_{\rm E})$.

The observational strategy is relatively unimportant, as long as there
are astrometric measurement on either side of a caustic.  Although it
is conceivable that observations might be scheduled at particularly
favorable times, such as times of maximum magnification or maximum
displacement of the image centroid, it is likely that the telescope
measuring the astrometric displacement will be oversubscribed.  A
simpler strategy would be to schedule periodic observations in
advance.  We have assumed that observations will be made every four
days, with a 24 hour delay after the event is recognized as a caustic
crossing binary, i.e., after the first caustic crossing.  That is, the
first observation comes $1-5$ days after the first caustic crossing.  
We assume observations are made for a total of 36 days, which corresponds
to approximately $2\te$ for the median event timescale.  Practically, 
observations should continue until after the second caustic crossing, and
the total number of observations should be at least as large as 
the number of parameters to be constrained. 

As we discuss in \S\ref{sec:errors} (see also Table 1), well covered binary lenses can be
fit photometrically with small errors on the parameters.  Thus, we
have assumed that all the parameters which can be fit from a single
photometric telescope are determined.  

Given a binary microlensing event from our ensemble, we use our
observational strategy to create a series of photometric and
astrometric measurements, respectively $A(t)$ and $\bt_{cl}(t)$
which we can combine into a single list of measurements $M_k$ each
with uncertainty $\sigma_k$.  Using the Fisher matrix technique, e.g., Gould \& Welch (1996), we
determine the covariance matrix $c_{ij}$ of the errors
\begin{equation}
c \equiv b^{-1}, \quad b_{ij} = \sum_k \sigma_k^{-2} \frac {\partial
M_k}{\partial a_i} \frac {\partial M_k}{\partial a_j} \, .
\end{equation}
Here the $a_i$ are the various parameters being fit.  The error in parameter $a_i$
is simply $\sigma_{a_i}=\sqrt{c_{ii}}$.

We assumed that the photometric uncertainty, $\sigma_k$, of the interferometric
telescope, is photon-noise-dominated,
and that the total telescope time, aperture, efficiency, filter width,
and source brightness are such that a total of $N=60,000$ photons would be
detected from an unmagnified source for a total photometric signal to
noise of 250.\footnote{Note that a telescope with an overall
efficiency of $\sim 30\%$ and
diameter $A_T$ collects $\sim 100(A_T/1{\rm m})^2$ photons per second
at $I=18$.}   We assume that the fractional 
photometric accuracy is simply $N^{-1/2}$, and that the
astrometric uncertainty is 
$\sigma_{\rm \theta}= N^{-1/2}\theta_f$, where $\theta_f$ is the width 
of the point spread function, or in the case of an interferometer, the fringe
separation.  Here we have assumed $\theta_f=2.5$mas.  It is trivial to scale our
results to brighter sources or larger telescopes: the fractional
uncertainty in $\thetae$ is simply proportional to $N^{-1/2}$.

We always fit for the four parameters required 
to establish a local astrometric frame, and $\thetae$ and $\alpha$,
the size scale and orientation of the microlensing excursion.  We also
assumed that the satellite which measures the astrometric motion is
0.2 AU from the (ground based) photometric measurements which fix the
lens parameters.  Thus, we can simultaneously fit for $\retilde$, the projected
Einstein ring radius, in
the manner of Graff \& Gould (2002).  In addition to this basic fit,
we also considered blending from luminous lenses and binary sources, which requires several additional
parameters.  We discuss parallax and blending 
in \S\S\ref{sec:parallax} and \ref{sec:lumlens}, respectively.

Our basic results are summarized in the leftmost curve in Figure 2.
We see that in the absence of blending, $\thetae$
can be determined with $<10\%$ uncertainty in 85\% of events.  The median error is $\sthe/\thetae \simeq 2.4\%$. This is comparable
to the uncertainty found by Gould \& Salim (1999) for single star events,
but with 50 times as many photons as we have assumed here.  Thus,
it is much easier to measure $\thetae$ for caustic crossing binary
lenses than for single lenses.

\subsection{Expected Fractional Errors on $\theta_*$}\label{sec:errors}

From the expression for $\theta_*$ (Eq.~\ref{eqn:thetas}), and assuming
the errors in $\thetae$, $\te$, $\dt$, and $\phi$ are small and
uncorrelated, the fractional error in $\theta_*$ is given by,
\begin{equation}
\frac{\sigma_{\theta_*}}{\theta_*} = 
\left[
\left(\frac{\sthe}{\thetae} \right)^2 + 
\left( \frac{\ste}{\te} \right)^2  +
\left( \frac{\sdt}{\dt}  \right)^2 +
\sphi^2 \cot^2\phi 
\right]^{1/2}
\label{eqn:errths}
\end{equation}
where $\sthe$, $\ste$, $\sdt$, and $\sphi$ are the uncertainties in
$\thetae$, $\te$, $\dt$ and $\phi$, respectively, and $\sphi$ is in
radians.  

To date, there exist in the published literature 10 microlensing
events for which the source star was well-resolved.  Unfortunately, for
four of the events, those presented in Alcock et~al.\ (2000a), no
estimate of the errors of the derived parameters is given.  We
therefore cannot use these events to explore the expected magnitudes
of $\ste$, $\sdt$, and $\sphi$.  

For the six events for which errors on the relevant fit parameters
were given or derivable, only two of them are generic binary-lens fold
caustic crossing events.  The errors $\ste$, $\sdt$, and $\sphi$ for
these two events, MACHO 98-SMC-1 and OGLE-1999-BUL-23, are presented
in Table 1.  For MACHO 98-SMC-1, these errors have been determined
from the ensemble of solutions presented in Albrow et~al.\ (1999b),
which were fits to the PLANET collaboration photometry, which only
covered the last half of the event, whereas the analysis of the
combined photometry of the EROS, MACHO, MPS, OGLE, and PLANET
collaborations (Afonso et~al.\ 2000) yields considerably smaller
errors.  We concentrate on the results of Albrow et~al.\ (1999b) here
in order to demonstrate the kinds of errors that result from
incomplete light curve coverage.  Note that $\sdt \sim 1\%$, which is
not surprising, since the accuracy with which $\dt$ can be determined
depends almost exclusively on the photometric coverage near the
caustic crossing.  However, the errors on the global parameters $\te$
and $\phi$ are quite large, $\ste \sim 20\%$ and $\sphi \sim 30\%$.
This is due to the fact that the PLANET photometry only covered the
latter half of the event, and contained no data prior to or during
the first caustic crossing.  Therefore
the global geometry was quite poorly constrained with their data
alone.  Such incomplete coverage would clearly jeopardize a precise
measurement of $\theta_*$.
This is in contrast to OGLE-1999-BUL-23, for
which both PLANET and OGLE obtained data before the first caustic
crossing.  In this case, it was possible to measure $\dt$, $\te$, and
$\phi$ to better than $1\%$ (Albrow et~al.\ 2001a).   

The remaining four include a single lens event (MACHO 95-BLG-30;
Alcock et al.\ 1997), a binary-lens event in which the source crossed
a cusp (MACHO 97-BLG-28; Albrow et~al.\ 1999a), a binary-lens event in
which the rotation of the binary was detected (MACHO 97-BLG-41; Albrow
et~al.\ 2000), and a binary-lens event for which both rotation and
parallax effects were detected (EROS BLG-2000-5; An et~al.\ 2002).  In general, the
previous discussion is not directly applicable to these events, as the
information in the light curves is not easily decomposed into the
parameters $\dt$, $\phi$ and $\te$.  Nevertheless, in all four cases the
dimensionless source size $\rho_*\equiv \theta_*/\thetae$ was
determined.  In three of the cases, $\delta \rho_*/\rho_* \la 1\%$.
In one case, MACHO 97-BLG-41, $\delta \rho_*/\rho_* \sim 5\%$ , due
primarily to the fact that there exists only a handful of data points
in which the source was resolved.  Thus, although it is difficult to
draw any general conclusions from these unique events, it is does seem
likely that errors of $\la 1\%$ are achievable for most types of
events in which the source is resolved.

Also presented in Table 1 are the determined values of 
the $I$-magnitude of the source, for all six
events.  At $I\sim 22$, the
source star for MACHO 98-SMC-1 would be too faint to target with most upcoming interferometers,
including SIM.  The other events are primarily bulge clump giants ($I\sim 15$), for
which accuracies of $\sthe/\thetae \la 5\%$ should be achievable with a reasonable
amount of exposure time with upcoming interferometers, and in particular with 
$\la 8$ minutes of SIM time (see \S\ref{sec:number} for a estimate of the required 
exposure times for SIM).  In these cases, the expected error
on $\theta_*$ is typically dominated by $\sthe$, and therefore we can
expect $\sigma_{\theta_*}/\theta_* \sim \sthe/\thetae \sim 5\%$.  The
source of OGLE-1999-BUL-23 is G/K subgiant ($R_* \sim 3R_{\odot}$),
and thus dimmer.  The time required to achieve an accuracy of
$\sthe/\thetae \sim 5\%$ will be larger by a factor of $\sim 16$, or $\sim 2~{\rm hr}$ for SIM.

\section{Discussion}\label{sec:discussion}

Our goal in \S\ref{sec:method} was to capture the essence of the
method of measuring $\theta_*$, and the discussions were therefore
somewhat oversimplified, and glossed over several important points.
In particular, we concentrated on fold caustic crossing binary-lens
events toward the bulge, whereas measurements of $\theta_*$ should be
possible in other, rarer, types of events, such as cusp-crossings and
single-lens events, and possibly events toward the Magellanic Clouds.
We also ignored various higher-order effects which could, in
principle, complicate the measurements.  We therefore briefly discuss
some of these complications and extensions.  We also discuss the
prospects for measuring the spectral type of the source, and also its
distance, in order to convert from angular radius $\theta_*$ to
physical radius $R_*$.  Finally, we present an example 
observing campaign aimed at measuring angular radii for a significant
sample of sources, outlining the resources required, and estimating
the number of $\theta_*$ measurements that might be made per year for
such a campaign.

\subsection{Single-lens, Cusp-crossing, and Magellanic Cloud Events}\label{sec:cusp}

Although we have focussed on fold caustic-crossing binary-lens events
toward the bulge, it is important to
emphasize that $\theta_*$ can, in principle, be measured for other
types of caustic-crossing events such as cusp-crossing events,
single-lens events, and all types of caustic-crossing events
toward the Magellanic clouds.  Indeed, in \S\ref{sec:errors} we discussed examples in the
literature of a single-lens and two cusp-crossing events for which a
$\sim 5\%$ measurement of $\theta_*$ would have been feasible.

In general, isolated cusp crossings, such as in MACHO 98-BLG-28,
cannot be predicted in advance, and thus planning for such events is
difficult, if not impossible.  However, one will still have some
advance warning of those cusp events which occur just after or in
place of second fold caustic crossings.  For such events, sufficient
photometric coverage of the crossing should routinely be possible.  In
all cases, it is more difficult to disentangle the information arising
from the cusp itself with the information from the global light curve.
This generally implies that the analysis of these light curves will be
more complicated, however this does not necessarily preclude an
accurate measurement of $\theta_*$.

Single-lens events are less desirable simply because they require a
factor of $\sim 50$ times more astrometric observing time to achieve
the same fractional accuracy in $\thetae$ as binary-lens events.
Since the astrometric observations are likely to be the most limited
resource, this makes single lens events considerably less attractive.

If it were possible to measure angular radii of stars in the
Magellanic clouds (MCs), this would be quite interesting, due to the
metal-poor nature of the stars.  Unfortunately, there are several
major hindrances to measuring $\theta_*$ for a substantial number of
stars in the MCs.  First, the event rates toward both the MCs are small,
and a large number of stars must be monitored just to detect a few
events per year.  Therefore, the number of caustic-crossing events is
quite low.  To date, there have been only two caustic crossing events
toward the MCs: MACHO 98-SMC-1, which we discussed in
\S\ref{sec:errors}, and MACHO LMC-9.  These events have source
magnitudes of $V_S = 22.4$ (Afonso et~al.\ 2000) and $V_S = 21.4$
(Alcock et al.\ 2000a), respectively, which brings up a second
difficulty: SIM cannot follow source stars fainter than $V\sim
20$, so these two events could not have been used to measure the
angular radii of their source stars.  In fact, even if the entire LMC
were monitored for microlensing, only $\sim 1$ event per year would
have $V_S \la 20$,
and this event would be from an evolved star. The probability of a caustic-crossing event
(either binary or single lens) is smaller by at least an order of
magnitude.  
The paucity of events and faintness of the source stars
might be circumvented if sufficiently rapid Target of Opportunity
times are available.  In this case, it might be possible to use
intrinsically fainter source stars, for which caustic-crossing event
will be more common, and measure the astrometric displacement during
the brief period of time when the source is highly magnified as it
crosses the caustic.  The maximum magnification of a source of
dimensionless size $\rho_*$ crossing a fold caustic is $A_{\rm max}
\sim \rho_*^{-1/2}$.  For main-sequence sources, $A_{\rm max} \ga 30$,
or more than three magnitudes, and thus sources with $V_S \la 23$ can
briefly be brightened to SIM detectability.  For example, the
source star of MACHO 98-SMC-1 was brighter than $V\sim 18$ for about 7
hours during the second caustic crossing.  Finally, even if the source
does attain a sufficient brightness to be measurable by SIM, it
remains to be seen whether the centroid varies sufficiently during
this time to provide an accurate measurement of $\thetae$.  This is
especially difficult in light of the fact that typical value of
$\thetae$ for self-lensing events toward the MCs are only an order of
magnitude larger than SIM's accuracy (Paczy\' nski 1998, Gould \& Salim 1999).  In summary, it appears
that it will be quite difficult to measure angular radii of stars in
the MCs using this method, especially if the majority of the events
seen toward these targets are due to self-lensing (Sahu 1994).

\subsection{Complications to $\thetae$ Measurement}\label{sec:complications}

The method we have presented here is only interesting if it can
feasibly be used to make precise $\theta_*$ measurements for a large
number of sources with reasonable expenditure of resources.  Since the
requisite astrometric instruments are likely to be the most limited
resource, it is crucial that accurate and unambiguous determinations
of $\thetae$ be generically possible using a few astrometric
measurements, when combined with the photometric light curve solution.
We have explained how a complete photometric solution {\it generally}
leads to a prediction for the astrometric centroid shift up to an
unknown scale $\thetae$ and orientation $\alpha$ on the sky.  However, this is true only 
under a number of
simplifying assumptions, including uniform motion of the observer,
source, and lens, dark lenses, isolated sources, and unique global
solutions.  If one or more of these assumptions are violated, then the
prediction for shape of the astrometric curve may not be unique, and
thus the measurement of $\thetae$ may be compromised.  We therefore
discuss each of these complications and under what conditions they may
be important.

\subsubsection{Binary Lens Degeneracies}\label{sec:degeneracies}

Binary lenses are characterized by two quantities: $q$, the mass
ratio, and $d$, the instantaneous projected separation in units of
$\thetae$.  It has been demonstrated both theoretically (Dominik 1999a)
and observationally (Afonso et~al.\ 2000, Albrow et~al.\ 2002) that
certain limiting cases of binary lenses can exhibit extremely similar
observable properties.  In particular, Dominik (1999) showed that the
binary-lens equation can be approximated by an single lens with
external shear, or Chang-Refsdal (CR) lens (Chang \& Refsdal
1979,1984), near the individual masses for widely-separated binaries
($d \gg 1$), and near the secondary (least massive) lens when $d \ll
1$.  Furthermore, near the center-of-mass of a close binary, the lens
equation is well-approximated by a quadrupole lens, and both
the quadrupole lens and CR-lens can exhibit extremely similar
magnifications when the quadrupole moment is equated to the shear (Albrow et~al.\ 2002).
Thus there can exist multiple degenerate solutions to an observed
photometric light curve, even with extremely accurate photometry.
However the astrometric behavior of
these degenerate solutions is very similar both in the
shape and overall scale of the astrometric curves, at least for the
close/wide degeneracy (Gould \& Han 2000).  Therefore this degeneracy should not affect the
determination of $\thetae$ using the prediction from the light curve.
It is likely that the other intrinsic denegeracies will also not
affect the determination of $\thetae$, since the degeneracy arises
from the lens equation itself, and thus affects both the photometric
and astrometric curve in the same manner.

Note that it is important that the normalization of $\thetae$ be
consistent for the two degenerate solutions.  For example, consider
the case of the close/wide degeneracy in MACHO 98-SMC-1 (Afonso et~al
2000, Gould \& Han 2000).  If one normalizes to the total mass of the
binary, the close solution implies a value of $\theta_{{\rm E},c} =
76\muas$, whereas the wide solution has $\theta_{{\rm E},w}=170\muas$.
Since the astrometric curves are essentially identical (both in shape
and scale), one might therefore suspect the inversion of this process
would yield two equally likely values of $\theta_*$ that differed by a
factor of $\sim 2$.  Of course, this `ambiguity' is wholly artificial,
and arises because the value of $\theta_{{\rm E},w}$ for the wide
binary is normalized to the entire mass of the binary, whereas the
lensing effects are basically caused by the least massive lens, since
$d=3.25$.  
Normalizing to the mass of the single lens, $\theta_{{\rm
E},w}'= \theta_{{\rm E},w} (1+q^{-1})^{-1/2}$, where here $q=0.24$,
and thus $\theta_{{\rm E},w}'=75 \muas$, essentially identical to the
close-binary solution.  Note that as $d$ approaches unity, the
identification of the `proper' $\thetae$ normalization 
becomes more nebulous, since the lenses can no longer be
considered independent.  However, the degeneracies also become less
severe as $d\rightarrow 1$.

Dominik (1999b) has also shown that poorly-sampled binary lens
light curve can also yield distinct degenerate solutions.  Note that
these solutions are `accidental' in the sense that they do not arise
from degeneracies in the lens equation itself.  Thus Han et al. (1999) 
found that such degenerate photometric light curves yield
astrometric curves which are widely different.  Such degeneracies
would prohibit the measurement of $\thetae$ using a few astrometric
measurements.  Therefore well-sampled photometric light curves are
essential for reliable measurements of $\theta_*$.

\subsubsection{Parallax}\label{sec:parallax}

If the two observers are displaced by a significant fraction of
$\tilde r_{\rm E}\equiv D\thetae$, the angular Einstein ring radius
projected onto the observer plane, then the source position $\bu$ as
seen by the two observers will be significantly different.  Since 
SIM will be in an Earth-trailing orbit, it will drift away from the
Earth at a rate of $\sim 0.1~{\rm AU~yr^{-1}}$.  Thus after 2.5 years
(half way through the SIM mission), it will be displaced from
the Earth by $\ell \sim 0.25~{\rm AU}$, which corresponds to a
displacement in the Einstein ring of,
\begin{equation}
|\delta \bu| = \frac{\ell}{\tilde r_{\rm E}} {| \sin \gamma|},
\label{eqn:simoff}
\end{equation}
where $\gamma$ is the angle between the the line of sight and the
Earth-SIM vector.  For typical bulge parameters,
\begin{equation}
\tilde r_{\rm E} = 7.6~{\rm AU} \left(\frac{M}{0.3 \msun}\right)^{-1/2},
\label{eqn:retilde}
\end{equation}
and therefore $|\delta \bu| \sim 3\%$. This implies that both the
magnification $A$ and the centroid shift $\dbtcl$ will be
significantly different (at a given time) as seen from the Earth and
SIM.  Since the value of $\tilde r_{\rm E}$ is not known {\it a
priori}, $|\delta \bu|$ cannot be predicted from the ground-based
photometry alone, and must be estimated from the astrometric data
itself.  Fortunately, SIM will likely have excellent
photometric capabilities (see Gould \& Salim 1999), and thus the
relative magnifications between the light curves from the ground and
SIM will provide additional constraints.   
Indeed, in our Monte Carlo simulations, we assumed that the astrometric
observer was displaced by $0.2~{\rm AU}$ from the Earth, and simultaneously fit
for both $\thetae$ and $\tilde r_{\rm E}$ (among other parameters), and
found that $\thetae$ could still be constrained quite accurately.  This
is because the information about $\tilde r_{\rm E}$ comes primarily
from the photometry, while the information about $\thetae$ comes primarily
from the astrometry.  Therefore, the two parameters are not degenerate.  
Note that a `byproduct' of
these measurements is a determination of the total mass of
the binary lens (Gould \& Salim 1999; Han \& Kim 2000; Graff \& Gould 2002)

There are two additional parallax effects.  One is due to the motion
of the Earth (or SIM) around the sun, and will become significant on
timescales that are a substantial fraction of a year, which
corresponds to many $\te$ for typical bulge events.  There is also a
second order effect that arises from the difference in projected
velocities between the Earth and SIM.  This effect is $\propto
v_{\oplus}/\tilde v $, where $\tilde v \equiv \tilde r_{\rm E}/\te$ is
the transverse velocity of the lens projected on the observer plane,
and is $\sim 800~{\rm km~s^{-1}}$ for typical bulge self-lensing
events.  Since the velocities and positions of the
Earth and SIM will be known, both of these effects can easily be included in the fit
for the microlensing parallax, and so do not present any additional difficulties.

\subsubsection{Luminous Lenses and Binary Sources}\label{sec:lumlens}

With its planned 10 meter baseline, SIM will have a resolution
of $\sim 10\mas$, sufficient to resolve the majority of unassociated
nearby stars that are blended with the source in ground-based
photometry (Han \& Kim 1999).  Since the photometric blending is
well-constrained in binary-lens events, unambiguous prediction of the
unblended astrometric behavior of the source is possible.  Thus,
blending will typically not affect the measurement of $\thetae$.
However, luminous lenses and companions to the source star with
separations $\la 10~\mas$ will not be automatically resolved by 
SIM (Jeong, Han \& Park 1999).  Dalal \& Griest (2001) have shown that, using two pointings,
this limit may be lowered to $\sim 3\mas$, however it is essentially
impossible to resolve multiple sources with separations below this
limit (e.g.\ binary source companions).  In these cases, all that will be measured is the total
centroid of all the sources in the resolution element.

The centroid in the presence
of luminous lenses, $\bphi_{cl,b}$, is related to the centroid
in the absence of blending, $\bphi_{cl}$, by
\begin{equation}
\bphi_{cl,b}=\left[\bphi_{cl} + \frac{f_t}{A}
\bphi_{b}\right] \left(1 + \frac{f_t}{A}\right)^{-1},
\label{eqn:dbpclt}
\end{equation}
where $f_t\equiv \sum_i F_{b,i}/F_0$ is the sum of the flux of all
unlensed sources (blends) relative to $F_0$, the baseline flux of the
lensed source,  $A$ is the magnification of the source, and
$\bphi_{b}$ is the centroid of the blends relative to the
origin of the lens.  From \eq{eqn:dbpclt}, 
it is clear blending is more complicated in astrometric microlensing than in
photometric microlensing: whereas photometric blending can be
described by one parameter, the blend fraction $f_t$,
astrometric blending requires two additional parameters, the centroid
of light of the blend $\bphi_{b}$.  A special case of astrometric blending is
bright lens blending.  In the single lens case, this eliminates the
blend location parameters, the location of the centroid of light is
the moving lens (i.e.\ $\bphi_{b}=0$).  In bright binary lens blending, only one parameter is
eliminated, the centroid of light is somewhere on the lens axis
between the two stars in the lens.  However, it will generally not be known {\it a priori} which 
case one is dealing with, and therefore $\bphi_{b}$ must be included as a parameter in the
astrometric fit. 

Blending is problematic because it effectively `dilutes' the astrometric shift
between two points in the lightcurve,
which is qualitatively similar to the effect of changing $\thetae$.
If the event is not well covered, these two effects can be quite
degenerate. In order to determine how degenerate blending is with the $\thetae$, we
have included in our Monte Carlo simulations a fixed amount of blended light of $f_t=1\%, 10\%$, and $90\%$.
We assume that $f_t$ is known (i.e.\ from photometry), but $\bphi_{b}$ is not.
The results are shown in Figure 2.  We find that, if
the blending fraction is close to 1, then the two effects are nearly
degenerate, and our fractional uncertainty in $\thetae$ increases by two orders
of magnitude to of order unity.  However, for $f_t\la 1\%$, the median error increases by less than
a factor of two.  In most cases, the blending will be known to be small from the photometric
data.  In these cases, the fractional uncertainty in $\thetae$ will not be seriously
degraded.  The few events with known large blending can be easily
jettisoned from the sample.

\subsubsection{Lens Rotation}\label{sec:rotation}

The photometric effects of lens rotation in binary microlensing events
has been explored theoretically by Dominik (1998) and has been
detected in event MACHO 1997-BUL-41 (Albrow et~al.\ 2000).  The
astrometric effects of rotating binary-lenses have not been explored,
and it is therefore difficult to draw any general conclusions as to
the importance of this effect.  However, to the extent that it is
detectable in the photometric light curve, lens rotation poses no
difficulties, as its astrometric effect should be predictable from the
global solution.  Effects that are photometrically undetectable but
astrometrically significant are potentially problematic.  

The amount that a binary-lens rotates during $\te$ is given by,
\begin{equation}
\psi\simeq 4.5^{\circ}d^{-3/2}\left(\frac{M}{0.3\msun}\right)^{1/4}\left(\frac{D'}{1.5\kpc}\right)^{-1/4}\left(\frac{v}{150\kms}\right)^{-1},
\label{eqn:binrot}
\end{equation}
where $D'=\dol\dls/\dos$, and 
assuming circular, face-on orbits.  Because the caustic cross-section
is maximized for binaries with separations of order $\thetae$, the
majority of detected caustic-crossings will have $d\sim 1$ (Baltz \& Gondolo 2001). 
Therefore, for typical events, the effects of binary rotation should be small if
astrometric observations are closely spaced with respect to the event
timescale $\te$.  This is generally also advantageous for the accurate recovery of $\thetae$ (See \S\ref{sec:thetae}).

\subsection{Measuring $\dos$}\label{sec:dos}

In this paper, we have emphasized the measurement of the
angular radius $\theta_*$, rather than the physical radius, $R_*$.  
However, it may also be interesting to measure $R_*$ for some events.
In order to do this, the distance to the source star must be measured independently. 
Fortunately, the astrometric accuracy needed to measure $\theta_*$ is generally
sufficient to measure the parallax $\pi_s$ of the source stars,
\begin{equation}
\pi_{s}=125\muas \left(\frac{\dos}{8\kpc}\right)^{-1}.
\label{eqn:pis}
\end{equation}
In order to measure $R_*$ to a similar accuracy as $\theta_*$ ($\sim 5\%$), $\pi_s$ must
be measured to somewhat better accuracy, which implies an astrometric error of $\sigma_\theta\la 5\muas$.
For SIM and an $I=18$ source, this is achievable with $\ga 7$ hours of integration, which is considerably
more time than is needed for the $\theta_*$ measurement alone.  However, it is important
to note that these measurements can be made after the microlensing event is over.  Therefore, it should
be possible to employ ground-based interferometers for the measurement of $\pi_s$, rather than 
spend precious SIM resources, at least for brighter sources.

\subsection{Typing the Source Star}\label{sec:source}

A measurement of $\theta_*$ is essentially useless if the
spectral type and luminosity class of the source is not known.  The
source stars of microlensing events can be typed in two ways.  The
first is to simply measure the color and apparent magnitude of the
source.  This information is generally acquired automatically from the
fit to the photometric data of the microlensing event.\footnote{Note
that blending in generally not a problem, as binary-lens events can
typically be easily deblended.}  By positioning the source star on a
color-magnitude diagram of other stars in the field, one can generally
type the source to reasonable accuracy.  The primary pitfalls of this
method are differential reddening and projection effects (i.e.\ the
source may in the foreground or background of the bulk of the stars in
the field).  

A more robust way of typing the source star is to acquire
spectra.  This is best done when the source is highly magnified as it
crosses a caustic, as this minimizes the effects of blended light
and increases the signal-to-noise.  Thus such measurements require
target-of-opportunity observations.  For highly-magnified events,
spectra with $S/N\sim 100$ per resolution element can be achieved with
exposure times of tens of minutes for low-resolution spectra (Lennon
et~al.\ 1996), or a couple of hours for high-resolution spectra
(Minniti et~al.\ 1998), using 8m or 10m-class telescopes.  Although
low-resolution spectra are sufficient for accurate 
\centerline{{\vbox{\epsfxsize=9.0cm\epsfbox{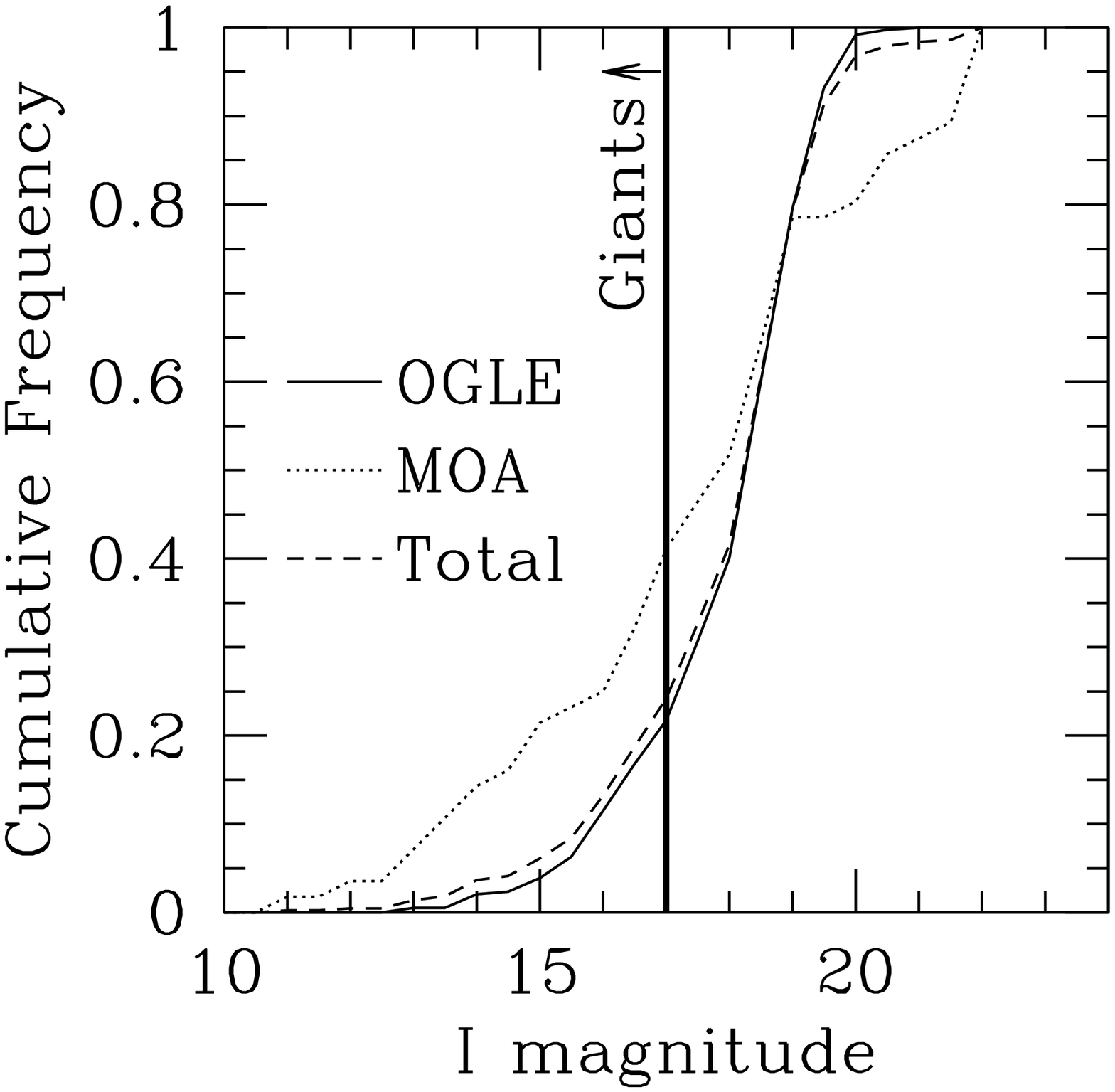}}}}
{\footnotesize { FIG. 3.--  The cumulative distribution of I magnitudes
for OGLE and MOA alerts in 2002.  The solid line is for OGLE,
the dotted line is for MOA, and the dashed line is the total.  The majority ($\sim 90\%$)
of alerts are from OGLE.
}
}\bigskip\\
spectral typing,
high resolution spectra are desirable for a number of other
applications, including resolution of the atmosphere of the source
star (Gaudi \& Gould 1999, Castro et~al.\ 2001, Albrow et~al.\ 2001b),
detailed abundance analysis (Minniti et~al.\ 1998), and detection of a
luminous lens (Mao, Reetz, \& Lennon 1998).  Note also that as a
byproduct, true space velocities of a sample of stars in the bulge will
be obtained by combining the proper motions and parallaxes of the sources acquired
from astrometric measurements with radial velocities determined from
the spectra.

\subsection{An Example Campaign}\label{sec:number}

In this section, we review the requirements for measuring $\theta_*$ for
the source stars of Galactic bulge microlensing events, and outline
the resources needed for a campaign aimed at measuring $\theta_*$ for a significant
number of sources.  

The first requirement is a large sample of caustic-crossing
binary-lens events from which to choose targets, which in turn
requires an even larger sample of microlensing events.  A large sample
is important in that it ensures that only interesting and promising
sources and events are followed.  Currently, both the OGLE and MOA
collaborations monitor many millions of stars in the Galactic bulge.
Both reduce their data real-time, enabling them to issue `alerts,'
notification of ongoing microlensing events (Udalski et~al.\ 1994,
Bond et~al.\ 2002a).\footnote{For online alerts, see
http://www.astrouw.edu.pl/$\sim$ftp/ogle/ogle3/ews/ews.html (OGLE) and http://www.roe.ac.uk/$\sim$iab/alert/alert.html (MOA).}  Combined,
these two collaborations should alert about $500$ events per year
(with the majority of alerts from OGLE).  Extrapolating from previous
results (Alcock et al.\ 2000a, Udalski et al.\ 2000), approximately
$5\%$ of these will be caustic-crossing binaries, or 25 events per
year.  Figure 3 shows the cumulative distribution of apparent (i.e.\
uncorrected for redenning) $I$ magnitudes of the 438 independent
 bulge microlensing alerts in 2002 for which baseline magnitudes were available.
Of these events, 382 ($87\%$) were alerted by OGLE, 61 ($14\%$) were alerted
by MOA, and 5 were alerted by both collaborations.  For the typical colors
of sources in the bulge ($V-I \sim 2$), only about $50\%$ of
the alerts have $V<20$, and thus would have been accessible to SIM.  The 
boundary between dwarfs and giants will occur at an apparent magnitude that  
depends on the color of the source, the distance to the source, and the reddening.
However, for definiteness we will simply assume that the
boundary between giants and main sequence stars occurs at roughly
$I=17$.  With this assumption, we can therefore expect that approximately $20\%$ of all alerted events
will be due to giant stars. Therefore, we can expect approximately
$20\%\times 5\% \sim 1\%$ of all alerts, or $\sim 5$ events (assuming 500 alerts), to be
caustic crossing events with giant sources, and $80\%\times 5\% \sim
4\%$, or $\sim 20$ events, to be caustic crossing events with main
sequence sources, $\sim 8$ of which will be bright enough to 
monitor with SIM. \footnote{This assumes that the detection efficiency
of binary-lens events does not depend on the $I$-magnitude of the
source.  In fact, deviations from the single-lens form will generally
be easier to detect in brighter sources, however this bias is likely
to be relatively small for caustic-crossing events, which generally
exhibit dramatic and easily-detectable deviations from the point lens
form.   This may seem in contradiction
with the fact that none of the five events toward the bulge
presented in Table 1 are main sequence sources.  However, this almost certainly a selection
effect: bright binary-lens events are currently preferentially monitored by
the follow-up collaborations, in order to achieve higher signal-to-noise during the second caustic crossing.}  
These numbers are likely to remain valid at least for the next
several years.  However, in the more distant future, and in particular
by the time SIM is launched, it is likely that the next generation of
microlensing survey collaborations will have come online.  Thus we can
expect that, when SIM time is operational,\footnote{The projected launch date
for SIM is currently 2009.} a considerably larger sample
of caustic-crossing events will be available.

Survey-type experiments are needed to discover microlensing events
toward the bulge, and survey-quality data is generally sufficient to
uncover the caustic-crossing nature of the target events.  However, as we
discussed in \S\ref{sec:dt}, more accurate and densely-sampled
photometry is generally needed during the caustic crossing in order to
measure $\dt$.  Currently, there are several collaborations with
dedicated (or substantial) access to 1-2m class telescopes distributed
throughout the southern hemisphere that closely monitor alerted
microlensing events with the goal of discovering deviations from the
single-lens form, with emphasis on search for extrasolar planets
(Albrow et~al.\ 1998, Rhie et al.\ 2000, Tsapras et al.\ 2001, Bond et
al.\ 2002b).  These collaborations have also been quite successful in
predicting and monitoring binary-lens caustic crossings.  It seems likely that
these collaborations, or similar ones, will still be in place when
the next generation of interferometers, or even SIM, come online.

In our Monte Carlo simulations we derived the expected precisions
$\sthe/\thetae$ assuming that the photometric errors were dominated by
photon statistics, and that a total of $N=60,000$ photons where collected
over the entire exposure time for each event.  This corresponds to
total exposure time of $T=1.6~{\rm hour}$ for SIM on an $I=18$ source.
We assume the fractional photometric uncertainty is $N^{-1/2}$, and
that the astrometric uncertainty was related to
the photometric uncertainty via the expression, $\sigma_{\rm \theta} =
N^{-1/2}\theta_f$, with $\theta_f = 2.5$mas, as appropriate
for SIM.  Since we assumed that the photometric and astrometric errors
are given simply by photon statistics, it is trivial to scale our
results for other total exposure times $T$ and source brightnesses
assuming the characteristics of SIM: $\sthe/\thetae \propto T^{-1/2}$,
and $\sthe/\thetae \propto 10^{0.2(I-18)}$.  For the purposes of
planning observations, and providing an order-of-magnitude estimate
for the number of angular radii that can be measured for a given
amount of SIM time, it is useful to derive an expression for the
exposure time required to achieve a given median photometric
precision.  To be conservative, we assume that the blending is small,
but non-negligible.  Specifically, we adopt the median error found for
the Monte Carlo simulations assuming a blend fraction of $f_t=1\%$,
which is $\sthe/\thetae \simeq 5.5\%$.  Then,
\begin{equation}
T\sim 2~{\rm hr} \left( \frac{\sthe/\thetae}{5\%} \right)^{-2}
10^{0.4(I-18)} .
\label{eqn:simtime}
\end{equation}
Thus, for giant sources ($I\simeq 15$), $7.4$ minutes are required
to achieve $5\%$ precision, whereas for main-sequence sources ($I
\simeq 20$), $12.3$ hours are required for $5\%$ precision, or $1.4$
hours for $15\%$ precision.

We have focussed here primarily on astrometric observations with SIM
because its capabilities are well-suited to this application.  The
basic requirements to be able to measure $\thetae$ accurately for the
events we have discussed are relatively high astrometric precisions,
$\sim 10~\muas$, and high sensitivity (via, i.e.\ large apertures), 
as the sources we are
considering are faint, $I=15-20$.  These faint sources are inaccessible
to current ground-based interferometers.  Upcoming
large-aperture, ground-based interferometers, such as the Very Large
Telescope Interferometer or the Keck Interferometer, should be able to
achieve the requisite astrometric precisions on all of the bright
($I\la 15$) giant events.  If the target microlensing source happens
to have a bright star within the isoplanatic angle, it may be possible
to employ phase referencing to extend sensitivity to very faint $(I\la
20)$ sources.  This would allow one to measure $\thetae$ for
main-sequences sources from the ground as well.  Finally, it may be
possible to determine $\thetae$ from single-epoch measurements of the
visibility and/or closure phase (Delplancke, G{\' o}rski, \& Richichi
2001, Dalal \& Lane 2003).  In this way, 
sensitivity could plausibly be extended to main-sequences sources by making carefully-timed 
interferometric measurements
of the source when it is highly-magnified during a caustic crossing.  However,
it is not clear if there exists enough structure in the image
positions during this time to extract $\thetae$.  This remains an interesting
topic for future study.   Non-targeted
space-based astrometric surveys, such as the Global Astrometric
Interferometer for Astrophysics, are generally not well-suited to this
application, due to the relatively sparse sampling of the source
stars.

Finally, access to target-of-opportunity time on 8-10m class
telescopes would allow for accurate spectral typing of the source
stars.  Several nights per bulge season would likely be adequate to
type the $\sim 25$ caustic-crossing events per year.  However, more time would be
required to perform some of the auxiliary science discussed in
\S\ref{sec:dos}, such as resolution of the source-star atmospheres.  

Thus, by combining alerts from survey collaborations, with
comprehensive ground-based photometry from follow-up collaborations with access
to dedicated (or semi-dedicated)
1m-class telescopes, and a modest allocation of a total 10 hours of SIM time,
it should be possible to measure the angular radii of $\sim 80$ giant
stars in the bulge to $5\%$, or $\sim 7$ main sequence stars to
$15\%$.  Several nights of target-of-opportunity time on $8-10$m
telescopes should allow for accurate spectral typing of the sources
via high or low-resolution spectroscopy.

\section{Conclusion}\label{sec:conclusion}

We have outlined a method to measure the angular radii $\theta_*$ of
giant and main sequence source stars of fold caustic-crossing binary
microlensing events toward the Galactic bulge.  Our method to measure
$\theta_*$ consists of four steps.  First, survey-quality data can be
used to discover and alert caustic-crossing binary-lensing events.
Such data is sufficient to characterize the event timescale $\te$ and
the angle $\phi$ of source trajectory with respect to the caustic.
Dense sampling of one of the caustic crossings yields the
caustic-crossing timescale $\Delta t$.  The global solution to the
binary-lens light curve yields a prediction for the trajectory of the
centroid of the source up to an unknown angle $\alpha$, and the scale,
$\thetae$.  Thus a few, precise astrometric measurements during
the course of the event yield $\thetae$.  The angular source radius
is then simply given by $\theta_*=\thetae(\dt/\te) \sin{\phi}$.

We argued, based on past experience with modeling binary-lens
events, that the parameters $\Delta t$, $\phi$, and $\te$ should be
measurable to a few percent accuracy, provided one caustic-crossing is
densely and accurately sampled, and the entire event is reasonably
well-covered.

We then performed a series of Monte Carlo experiments that
demonstrated that astrometric measurements during the course of the
binary-lens event should allow for the determination of $\thetae$ to
$\sim2\%$ accuracy, assuming photon-limited statistics and a total of
60,000 photons per event.  This is a factor of $\sim 50$ fewer
photons than are required to measure $\thetae$ to the same precision
in single-lens events and corresponds to an
exposure time of $T=1.6$~hour with SIM on an $I=18$ source.
Therefore, it should be possible to measure $\theta_*$ for a
significant sample of giant and main-sequence stars in the bulge with
reasonable expenditure of resources.

\bigskip
We would like to thank Neal Dalal for helpful conversations.  We would
also like to thank the anonymous referee for useful comments and suggestions.
This work was supported by NASA through a Hubble Fellowship grant
from the Space Telescope Science Institute, which is operated by the
Association of Universities for Research in Astronomy, Inc., under
NASA contract NAS5-26555, by JPL contract 1226901, and by the Science
Research Center (SRC) of the Korean Science and Engineering Foundation (KOSEF).

{}

\end{document}